\documentclass[aps,prl,superscriptaddress,showpacs,floatfix,twocolumn]{revtex4}

  \usepackage[dvips]{graphicx}

\begin{document}

\title{Centrality Dependence of Charm Production from Single Electrons 
       Measurement in Au+Au Collisions at $\sqrt{s_{NN}} = 200$\,GeV}

\newcommand{\abilene}{Abilene Christian University, Abilene, TX 79699, USA}
\newcommand{\acadsin}{Institute of Physics, Academia Sinica, Taipei 11529, Taiwan}
\newcommand{\banaras}{Department of Physics, Banaras Hindu University, Varanasi 221005, India}
\newcommand{\barc}{Bhabha Atomic Research Centre, Bombay 400 085, India}
\newcommand{\bnl}{Brookhaven National Laboratory, Upton, NY 11973-5000, USA}
\newcommand{\caucr}{University of California - Riverside, Riverside, CA 92521, USA}
\newcommand{\ciae}{China Institute of Atomic Energy (CIAE), Beijing, People's Republic of China}
\newcommand{\cns}{Center for Nuclear Study, Graduate School of Science, University of Tokyo, 7-3-1 Hongo, Bunkyo, Tokyo 113-0033, Japan}
\newcommand{\columbia}{Columbia University, New York, NY 10027 and Nevis Laboratories, Irvington, NY 10533, USA}
\newcommand{\dapnia}{Dapnia, CEA Saclay, F-91191, Gif-sur-Yvette, France}
\newcommand{\debrecen}{Debrecen University, H-4010 Debrecen, Egyetem t{\'e}r 1, Hungary}
\newcommand{\fsu}{Florida State University, Tallahassee, FL 32306, USA}
\newcommand{\gsu}{Georgia State University, Atlanta, GA 30303, USA}
\newcommand{\hiroshima}{Hiroshima University, Kagamiyama, Higashi-Hiroshima 739-8526, Japan}
\newcommand{\ihepprot}{Institute for High Energy Physics (IHEP), Protvino, Russia}
\newcommand{\isu}{Iowa State University, Ames, IA 50011, USA}
\newcommand{\jinrdubna}{Joint Institute for Nuclear Research, 141980 Dubna, Moscow Region, Russia}
\newcommand{\kaeri}{KAERI, Cyclotron Application Laboratory, Seoul, South Korea}
\newcommand{\kangnung}{Kangnung National University, Kangnung 210-702, South Korea}
\newcommand{\kek}{KEK, High Energy Accelerator Research Organization, Tsukuba-shi, Ibaraki-ken 305-0801, Japan}
\newcommand{\kfki}{KFKI Research Institute for Particle and Nuclear Physics (RMKI), H-1525 Budapest 114, POBox 49, Hungary}
\newcommand{\korea}{Korea University, Seoul, 136-701, Korea}
\newcommand{\kurchatov}{Russian Research Center ``Kurchatov Institute", Moscow, Russia}
\newcommand{\kyoto}{Kyoto University, Kyoto 606, Japan}
\newcommand{\labllr}{Laboratoire Leprince-Ringuet, Ecole Polytechnique, CNRS-IN2P3, Route de Saclay, F-91128, Palaiseau, France}
\newcommand{\lawllnl}{Lawrence Livermore National Laboratory, Livermore, CA 94550, USA}
\newcommand{\losalamos}{Los Alamos National Laboratory, Los Alamos, NM 87545, USA}
\newcommand{\lpc}{LPC, Universit{\'e} Blaise Pascal, CNRS-IN2P3, Clermont-Fd, 63177 Aubiere Cedex, France}
\newcommand{\lund}{Department of Physics, Lund University, Box 118, SE-221 00 Lund, Sweden}
\newcommand{\muenster}{Institut f\"ur Kernphysik, University of Muenster, D-48149 Muenster, Germany}
\newcommand{\myongji}{Myongji University, Yongin, Kyonggido 449-728, Korea}
\newcommand{\nagasaki}{Nagasaki Institute of Applied Science, Nagasaki-shi, Nagasaki 851-0193, Japan}
\newcommand{\newmex}{University of New Mexico, Albuquerque, NM 87131, USA}
\newcommand{\nmsu}{New Mexico State University, Las Cruces, NM 88003, USA}
\newcommand{\ornl}{Oak Ridge National Laboratory, Oak Ridge, TN 37831, USA}
\newcommand{\orsay}{IPN-Orsay, Universite Paris Sud, CNRS-IN2P3, BP1, F-91406, Orsay, France}
\newcommand{\pnpi}{PNPI, Petersburg Nuclear Physics Institute, Gatchina, Russia}
\newcommand{\riken}{RIKEN (The Institute of Physical and Chemical Research), Wako, Saitama 351-0198, JAPAN}
\newcommand{\rkrbrc}{RIKEN BNL Research Center, Brookhaven National Laboratory, Upton, NY 11973-5000, USA}
\newcommand{\saispbstu}{St. Petersburg State Technical University, St. Petersburg, Russia}
\newcommand{\saopaulo}{Universidade de S{\~a}o Paulo, Instituto de F\'{\i}sica, Caixa Postal 66318, S{\~a}o Paulo CEP05315-970, Brazil}
\newcommand{\seoulnat}{System Electronics Laboratory, Seoul National University, Seoul, South Korea}
\newcommand{\stonybrkc}{Chemistry Department, Stony Brook University, SUNY, Stony Brook, NY 11794-3400, USA}
\newcommand{\stonycrkp}{Department of Physics and Astronomy, Stony Brook University, SUNY, Stony Brook, NY 11794, USA}
\newcommand{\subatech}{SUBATECH (Ecole des Mines de Nantes, CNRS-IN2P3, Universit{\'e} de Nantes) BP 20722 - 44307, Nantes, France}
\newcommand{\tenn}{University of Tennessee, Knoxville, TN 37996, USA}
\newcommand{\titech}{Department of Physics, Tokyo Institute of Technology, Tokyo, 152-8551, Japan}
\newcommand{\tsukuba}{Institute of Physics, University of Tsukuba, Tsukuba, Ibaraki 305, Japan}
\newcommand{\vandy}{Vanderbilt University, Nashville, TN 37235, USA}
\newcommand{\waseda}{Waseda University, Advanced Research Institute for Science and Engineering, 17 Kikui-cho, Shinjuku-ku, Tokyo 162-0044, Japan}
\newcommand{\weizmann}{Weizmann Institute, Rehovot 76100, Israel}
\newcommand{\yonsei}{Yonsei University, IPAP, Seoul 120-749, Korea}
\affiliation{\abilene}
\affiliation{\acadsin}
\affiliation{\banaras}
\affiliation{\barc}
\affiliation{\bnl}
\affiliation{\caucr}
\affiliation{\ciae}
\affiliation{\cns}
\affiliation{\columbia}
\affiliation{\dapnia}
\affiliation{\debrecen}
\affiliation{\fsu}
\affiliation{\gsu}
\affiliation{\hiroshima}
\affiliation{\ihepprot}
\affiliation{\isu}
\affiliation{\jinrdubna}
\affiliation{\kaeri}
\affiliation{\kangnung}
\affiliation{\kek}
\affiliation{\kfki}
\affiliation{\korea}
\affiliation{\kurchatov}
\affiliation{\kyoto}
\affiliation{\labllr}
\affiliation{\lawllnl}
\affiliation{\losalamos}
\affiliation{\lpc}
\affiliation{\lund}
\affiliation{\muenster}
\affiliation{\myongji}
\affiliation{\nagasaki}
\affiliation{\newmex}
\affiliation{\nmsu}
\affiliation{\ornl}
\affiliation{\orsay}
\affiliation{\pnpi}
\affiliation{\riken}
\affiliation{\rkrbrc}
\affiliation{\saispbstu}
\affiliation{\saopaulo}
\affiliation{\seoulnat}
\affiliation{\stonybrkc}
\affiliation{\stonycrkp}
\affiliation{\subatech}
\affiliation{\tenn}
\affiliation{\titech}
\affiliation{\tsukuba}
\affiliation{\vandy}
\affiliation{\waseda}
\affiliation{\weizmann}
\affiliation{\yonsei}
\author{S.S.~Adler}	\affiliation{\bnl}
\author{S.~Afanasiev}	\affiliation{\jinrdubna}
\author{C.~Aidala}	\affiliation{\bnl}
\author{N.N.~Ajitanand}	\affiliation{\stonybrkc}
\author{Y.~Akiba}	\affiliation{\kek} \affiliation{\riken}
\author{J.~Alexander}	\affiliation{\stonybrkc}
\author{R.~Amirikas}	\affiliation{\fsu}
\author{L.~Aphecetche}	\affiliation{\subatech}
\author{S.H.~Aronson}	\affiliation{\bnl}
\author{R.~Averbeck}	\affiliation{\stonycrkp}
% ...
\collaboration{PHENIX Collaboration} \noaffiliation

\date{\today}

\begin{abstract}

The PHENIX experiment has measured mid-rapidity transverse momentum
spectra ($0.4 < p_T < 4.0$\,GeV/$c$) of single electrons as a function
of centrality in Au+Au collisions at $\sqrt{s_{NN}} = 200$\,GeV.
Contributions to the raw spectra from photon conversions and Dalitz
decays of light neutral mesons are measured by introducing a thin
(1.7\% $X_0$) converter into the PHENIX acceptance and are
statistically removed. The subtracted ``non-photonic'' electron
spectra are primarily due to the semi-leptonic decays of hadrons
containing heavy quarks (charm and bottom).
For all centralities, 
charm production cross section is found to scale with the nuclear
overlap function, $T_{AA}$. For minimum-bias collisions
the charm cross section per binary collision is 
$N_{c\overline{c}}/T_{AA} = 622 \pm 57 {\rm \,(stat.)} \pm 160 {\rm
\,(sys.)} \,\mu$b.
\end{abstract}

\maketitle



In central Au+Au collisions at $\sqrt{s_{NN}} = 200$\,GeV $\pi^0$'s
and charged hadrons are strongly suppressed at high transverse
momentum ($p_T$)~\cite{PhenixAuAu200GeVPi0, 
StarAuAu130GeV200GeVChargedHighpT, PhenixAuAu200GeVChargedHighpT}. In
contrast, a modest high-$p_T$ enhancement is observed in d+Au
collisions at the same energy~\cite{PhenixdAu200GeVPi0, 
StardAu200GeVChargedHighPt}. This strongly suggests that the
suppression observed in Au+Au collisions is caused by final-state
effects ({\it e.g.}, parton energy loss in a dense medium produced
in the reaction~\cite{Mustafa:1997pm,Lin:1997cn}).

Heavy quarks (charm and bottom) are 
complementary probes of the hot and dense matter produced in high 
energy heavy ion collisions.  
Due to their large masses, charm and bottom cross sections are
calculable via pQCD
and their yield is sensitive to the initial gluon density~\cite{Appel}.
It has been predicted that heavy quarks suffer less energy 
loss than light quarks while traversing partonic matter due to
 the 
``dead cone'' effect~\cite{Deadcone,ElossMig,ElossZhang}.  
This can be studied
through systematic measurements of the $p_T$ spectra of open heavy
flavor. In addition, the open-charm yield is an important baseline for
understanding $J/\psi$ production which has been predicted 
to be either suppressed~\cite{Matsui} or enhanced~\cite{JpsiEnhance} 
in the presence of deconfined quarks and gluons.

The PHENIX experiment observed that inclusive single electrons 
in central and minimum-bias Au+Au
collisions at $\sqrt{s_{NN}} = 130$\,GeV were 
produced in
excess of purely ``photonic" contributions
(primarily due to $\pi^0$ Dalitz decays
and conversion of $\pi^0$ photons in the detector material)
\cite{Ppg11}.  This excess is consistent with the expected charm
production, assuming that it scales with the number of binary
nucleon-nucleon collisions ($N_{coll}$), or equivalently, with the
nuclear overlap function, $T_{AA}$.  In this Letter, we present results
on the 
single electron measurement 
in $\sqrt{s_{NN}} = 200$\,GeV Au+Au collisions.
The new data have higher statistics and smaller
systematic errors than the 130\,GeV data, allowing us to measure charm
production as a function of collision centrality.



The data used in this analysis were collected 
by the PHENIX experiment~\cite{PhenixDetector} 
during the 2001 run period of the Relativistic Heavy Ion Collider. 
A coincidence of the beam-beam
counters (BBC) and the zero degree calorimeters (ZDC) provides the
minimum-bias trigger ($92.2^{+2.5}_{-3.0}$\,\% of the 6.8 $\pm$ 0.5 barn 
Au+Au inelastic cross section). 
The centrality is determined by the correlation
between the multiplicity 
measured by the BBC
and the energy of spectator neutrons measured by the ZDC.  The BBC
also measures the collision vertex, $z$, with resolution $\sigma =
0.7$\,cm. Events are required to have $|z| < 20$\,cm to
eliminate electrons originating from the central magnet.

Charged particles are measured by the PHENIX east-arm spectrometer
($|\eta|<0.35$, $\Delta\phi = \pi/4$) with resolution 
$\sigma_p/p \simeq 0.7\%
\oplus 1.0\%\:p ({\rm GeV}/c)$.  Tracks are reconstructed with the drift
chamber (DC) and the first layer of pad chambers (PC1) and confirmed
by requiring an electromagnetic calorimeter (EMC) \cite{PHENIXEMC}
matching hit within 2 standard deviations in position.
Electron
candidates are required to have at least three associated 
hits in the ring imaging \v{C}erenkov detector (RICH)
that pass a ring shape cut, and are required
to pass a timing cut in either the EMC or the time-of-flight detector.
After these cuts,
a clear electron signal is observed as a
narrow peak at $E/p = 1$. By requiring $-2\sigma < (E-p)/p < 3\sigma$, 
background
from hadrons, which deposit only a fraction of their energy in the
EMC, and non-vertex electrons, which have mis-reconstructed
momenta, is further reduced.  Remaining background in the
electron sample, due to accidental coincidences between RICH hits and
hadron tracks, is estimated ($\approx 10\%$) and subtracted by an
event-mixing method.



Inclusive electrons contain two components: 
(1) ``non-photonic'' -- primarily
semi-leptonic decays of mesons containing heavy (charm and bottom)
quarks, and
(2)``photonic'' -- Dalitz decays of light neutral mesons
($\pi^0$, $\eta$, $\eta'$, $\rho$, $\omega$ and $\phi$) and photon conversions in
the detector material. 
To separate these two components, a photon converter (a
 thin brass tube of 1.7\% radiation length surrounding the beam pipe at 
 $r = 29$\,cm) was installed. 

 \begin{figure}[htbp]
 \includegraphics[width=1.0\linewidth]{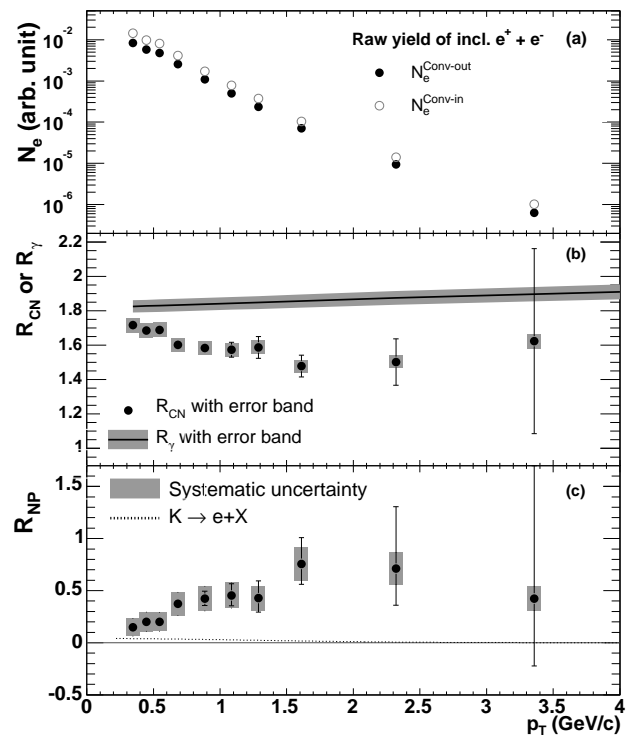}
 \caption{\label{fig:conv_method} Shown vs $p_T$ (a) Raw $e^\pm$ spectra measured with 
  the converter in (open circles) and out (closed circles).  (b) Ratio of 
  the converter in/out $e^\pm$ yields ($R_{CN}$, points) and ratio of photonic 
  $e^\pm$ yield with/without the converter ($R_{\gamma}$, line and shaded band). 
  (c) Ratio of non-photonic to photonic $e^\pm$ yields 
  ($R_{NP}$, points) and contribution from kaon decays (dashed line).
 }
 \end{figure}

We analyzed 2.2\,M (2.5\,M) events with the converter in (out). 
The corresponding raw electron $p_T$ spectra
for minimum-bias collisions are shown in 
Fig.~\ref{fig:conv_method}(a). 
The photon converter multiplies the photonic contribution
to the electron yield by a factor $R_{\gamma}$:
\vspace{-2mm}
\begin{eqnarray}
  N_{e}^{Conv-out}  =  &           N_{e}^{\gamma} &+ \; N_{e}^{non-\gamma} \label{eq:convout} \\
  N_{e}^{Conv-in}   =  & R_{\gamma}N_{e}^{\gamma} &+ \; (1-\epsilon) \cdot N_{e}^{non-\gamma} \label{eq:convin}
\vspace{-2mm}
\end{eqnarray}
Here $N_{e}^{Conv-in}$ ($N_{e}^{Conv-out}$) is the measured electron
yield with (without) the converter; $N_{e}^{\gamma}$
($N_{e}^{non-\gamma}$) is the electron yield due to
the photonic (non-photonic) component; and $\epsilon$($\approx$ 2.1\%) 
represents a small loss of electrons due to the converter.
We next define $R_{CN}$ as the ratio of the raw electron yield with and
without the converter. Dividing Eq. (\ref{eq:convin}) by
Eq. (\ref{eq:convout}) and defining $R_{NP} \equiv
N_e^{non-\gamma}/N_e^{\gamma}$, one has:
\vspace{-2mm}
\begin{equation}
  R_{CN} \equiv \frac{N_e^{Conv-in}}{N_e^{Conv-out}} = \frac{ R_{\gamma} + (1-\epsilon)R_{NP}}{1 + R_{NP}} 
\label{eq:R_CN}
\vspace{-2mm}
\end{equation}
If there were no contribution from non-photonic component ($R_{NP}$ = 0),
then $R_{CN} = R_{\gamma}$.

The photonic electron yield per photon is approximately given by $Y
\propto \delta + \frac{7}{9}t$, where 
$\delta$ is the Dalitz branching ratio per $\gamma$ relative to 2$\gamma$ 
(for $\pi^0, \eta , \eta^{\prime}$) or 1$\gamma$ (for $\rho$, $\omega$ 
and $\phi$) decay, and $t$ is the thickness of the conversion material 
in radiation length ($X_0$).
The factor $\frac{7}{9}$ is the
approximate probability for a $\gamma$ to convert in one $X_0$.
Plugging in $\delta^{\pi^0} =
0.6\%$, $t \approx 1.1\%$ ($t \approx 2.8\%$) for converter in (out) we
find $R_{\gamma}^{\pi^0} = Y^{Conv-in} / Y^{Conv-out} \approx 1.9$.  
There is some $p_T$ dependence in
the complete formula for $Y$ and the value of $\delta$ is
species-dependent ($\delta^{\eta} \approx 0.8\%$), so we perform a full
GEANT~\cite{Geant} simulation with and without the converter to calculate
$R_{\gamma}$. We determine $R_{\gamma}$ for $\pi^0$ and $\eta$
separately. We use the $\pi^0$ spectrum measured by PHENIX 
~\cite{PhenixAuAu200GeVPi0} as the
input for the $\pi^0$ simulation and assume $m_T$ scaling ($p_T
\rightarrow \sqrt{p_T^2 + M_{\eta}^2 - M_{\pi}^2}$, normalized at high
$p_T$ to $\eta/\pi^0 = 0.45 \pm 0.1$, 
based on the world data of $\eta/\pi^0$ ratio) to obtain the input
for the $\eta$ simulation. Contributions from other mesons which undergo
Dalitz decay ($\eta^{\prime}, \rho, \omega, \phi$) are small (6\% at $p_T =
3$\,GeV/$c$, and smaller at lower $p_T$). Since they have $\delta
\approx \delta^{\eta}$ we assign them $R_{\gamma} = R_{\gamma}^{\eta}$.
When calculating the combined $R_{\gamma}$ we use the particle ratios at
high $p_T$ ($\eta^{\prime}/\pi^{0} = 0.25 \pm 0.13, 
\rho/\pi^0 = \omega/\pi^{0} = 1 \pm 0.5, \phi/\pi^{0} = 0.4 \pm 0.2$). 
The $\phi/\pi^{0}$ ratio used here is consistent with our $\pi^{0}$ and 
$\phi$ measurement~\cite{PhenixPhi}.
The uncertainties in the particle ratios are included in the systematic 
uncertainties of $R_{\gamma}$.
For this method it is essential 
that the amount of material
is accurately modeled in the simulation. We compared the yield of
identified photon conversion pairs in the data and in the
simulation and conclude that the simulation reproduces $R_{\gamma}$
within $\pm 2.0\%$. 
This uncertainty is included in the overall systematic uncertainty.


Fig.~\ref{fig:conv_method}(b) shows $R_{CN}$ and
$R_{\gamma}$ vs. $p_T$. $R_{CN}$ gradually decreases with increasing
$p_T$, while $R_{\gamma}$ slightly increases with $p_T$. 
The difference between $R_{CN}$ and $R_{\gamma}$ indicates 
the existence of non-photonic electrons. 
Fig.~\ref{fig:conv_method}(c) shows $R_{NP}$ obtained from $R_{\gamma}$
and $R_{CN}$ using Eq. (\ref{eq:R_CN}). $R_{NP}$ increases with $p_T$ and
is more than 30\% for $p_T > 0.6$~GeV/$c$.
The small amount of conversion material in the PHENIX 
detector allows a sensitive measurement of $R_{CN}$.


Background from kaon decays ($K \rightarrow \pi e \nu$) and
di-electron decays of $\rho$, $\omega$ and $\phi$ remain in the
non-photonic electron yield. The background from kaon decays is
estimated with a GEANT simulation using the kaon $p_T$ spectrum
measured by PHENIX~\cite{Pidspectra} as input.  The contribution
of kaon decays to the non-photonic yield, shown in
Fig.~\ref{fig:conv_method}(c), is 18\% at $p_T =
0.4$\,GeV/$c$ and decreases rapidly to less than 6\% for $p_T >
1$\,GeV/$c$.

To calculate background from the $e^+e^- $ decays of $\rho$, $\omega$ 
and $\phi$, 
we first generate spectra by applying $m_T$ scaling to the PHENIX $\pi^0$
spectrum, as described above. 
The contribution of these decays to the non-photonic electrons is
$<3$\% for all $p_T$. 
Background from $J/\psi \rightarrow e^+e^-$ decays and 
from Drell-Yan pairs is negligible.
Possible enhancement of low mass di-leptons through
$\pi+\pi \rightarrow \rho \rightarrow e^+e^-$, as reported in Pb+Pb
collisions at the SPS~\cite{CERES}, would contribute to the
non-photonic electrons. However, this is neglected since the estimated
$\rho$ contribution in the absence of enhancement is only $\approx
0.6\%$ over all $p_T$.

\begin{figure}[htbp]
 \includegraphics[width=1.0\linewidth]{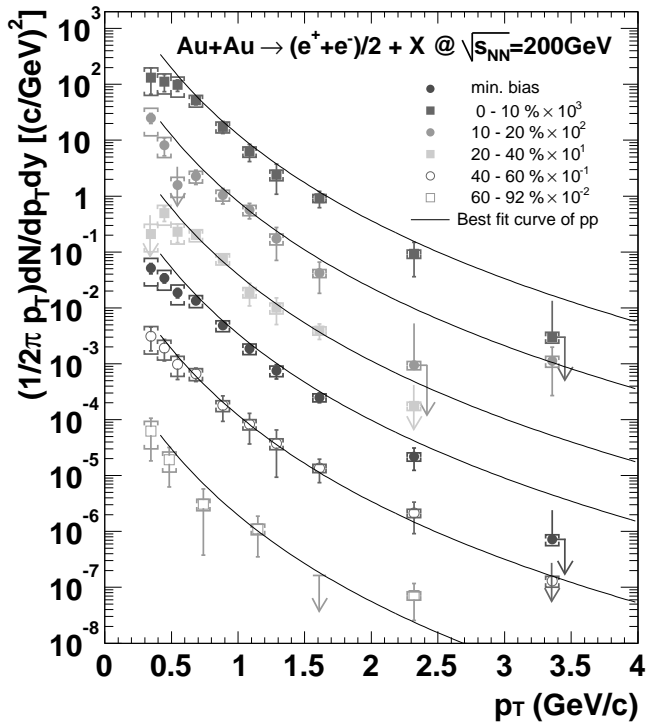}
 \caption{\label{fig:correct_spectra} Fully corrected heavy flavor 
 electron
 $p_T$ spectra for different Au+Au centralities scaled by successive
 factors of ten for clarity. Error bars (brackets) correspond to
 statistical (systematic) uncertainties.  Curves are described in the
 text.}
 \end{figure}

After these backgrounds are subtracted the only other significant
source of non-photonic electrons is the semi-leptonic decay of
heavy flavor (charm and bottom). The raw spectrum of heavy flavor
electrons is corrected for geometrical acceptance ($\epsilon_{geo}$),  
track reconstruction efficiency ($\epsilon_{rec}$) and electron 
identification efficiency ($\epsilon_{eID}$) 
determined by GEANT simulation. 
The efficiency 
$\epsilon_{geo} \times \epsilon_{rec}$ is about 11\% of $dN_e/dy$, 
and $\epsilon_{eID}$ is about 65\% as confirmed with 
electrons identified through photon conversion. 
Correction of multiplicity dependent efficiency losses, estimated by
embedding simulated electron tracks into real events, is
$p_T$-independent and increases from 5\% to 26\% from peripheral to
central collisions. The $1\sigma$ systematic uncertainty of these 
corrections is 11.8\%.
Fully corrected heavy flavor electron spectra are shown in
Fig.~\ref{fig:correct_spectra} for minimum-bias collisions and for
five centrality bins. 

PHENIX has also measured the heavy flavor electron spectrum in $p+p$
collisions at $\sqrt{s_{NN}} = 200$\,GeV~\cite{Ppelectron}.  The lines
in Fig.~\ref{fig:correct_spectra} show the best fit curve of this
spectrum, scaled by $T_{AA}$ for each Au+Au centrality bin.  Here,
$T_{AA}$ is the nuclear overlap function calculated by a Glauber
model~\cite{PhenixAuAu200GeVPi0}(Table \ref{tab:cent_TAA}).  
The Au+Au data points
are in reasonable agreement with the $p+p$ fit in all centrality bins.


\begin{figure}[htbp]
 \includegraphics[width=1.0\linewidth]{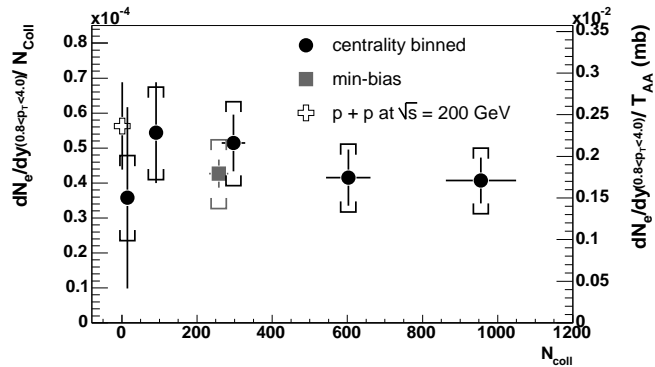}
 \caption{\label{fig:dndy} Non-photonic electron yield 
  ($0.8 < p_T < 4.0$\,GeV/$c$) 
  measured in Au+Au reactions at 200 GeV scaled by $N_{coll}$ 
  (left-hand scale) as a function of centrality given by $N_{coll}$.
  This electron yield translates to the electron cross section per
  $NN$ collision in the above $p_T$ range (right-hand scale).
  The yield in $p+p$ collisions at 200 GeV is also shown 
  \cite{Ppelectron}.
 }
 \end{figure}

To quantify the centrality dependence of heavy flavor production, 
we calculated the integrated yield $dN_e/dy$ ($0.8 < p_T < 4.0$\,GeV/$c$) 
and fit it to $A N_{coll}^{\alpha}$, 
where $\alpha = 1$ is the expectation in the absence of medium effects. 
In this comparison, most of the systematic effects will cancel.
Figure~\ref{fig:dndy} shows
$dN_e/dy(0.8<p_T<4.0)/N_{coll} $ vs. $N_{coll}$ for
minimum-bias and five centrality bins in Au+Au collisions
and $p+p$ collisions.
We find $\alpha=0.938 \pm 0.075$(stat.)$ \pm 0.018$(sys.).
If $p+p$ data is included, $\alpha=0.958 \pm 0.035$(stat.).
This shows that
the total yield of heavy flavor electrons for all centralities is
consistent with $N_{coll}$ scaling. 


For each centrality bin we scale the heavy flavor electron spectrum
($p_T > 0.8$\,GeV/$c$) by $T_{AA}$ and fit it with a PYTHIA
calculation of the electron spectrum resulting from leading order
charm and bottom production. 
We used PYTHIA 6.205 with a modified set of parameters
(described in~\cite{Ppg11}) and CTEQ5L PDFs~\cite{CTEQ}. Based on
experimental input ~\cite{CDF,PDGCharm}
we modified the PYTHIA default charm ratios,
using instead $D^+/D^0 = 0.45 \pm 0.1$, $D_s/D^0 = 0.25 \pm 0.1$,
$\Lambda_c/D^0 = 0.1 \pm 0.05$. 
This gives a $c$-quark $\rightarrow e$ total branching ratio of 
$9.5 \pm 0.4\%$. 
The scaled charm and bottom cross
sections are treated as fit parameters, although we find that our data
are restricted to $p_T$ values which are only sensitive to charm
production. We evaluated the systematic error due to background
subtraction 
($\approx 21\%$) 
by refitting to the electron spectrum at
the minimum and maximum of its $1\sigma$ systematic error band. 
The change of the $p_T$ range for fitting the non-photonic electron 
spectrum gives 3\% systematic error for minimum-bias collisions.
The systematic error due to the PYTHIA spectral shape ($\approx 11\%$) is
dominated by the uncertainty in $\langle k_T \rangle = 1.5 \pm 0.5$\,GeV/$c$. 
Different PDFs yield a systematic error of 6.2\% for the rapidity-integrated
cross section. 
Systematic errors in $T_{AA}$ are tabulated in \cite{PhenixAuAu200GeVPi0}. 
These systematic errors are added in quadrature to give the
overall systematic error on the charm cross section. For minimum-bias
collisions we obtain
$\frac{1}{T_{AA}}\frac{dN_{c\overline{c}}}{dy}|_{y=0} = 143 \pm 13
{\rm (stat.)} \pm 36 {\rm (sys.)}\,\mu$b and $N_{c\overline{c}}/T_{AA}
= 622 \pm 57 {\rm (stat.)} \pm 160 {\rm (sys.)}\,\mu$b.  Results for
all centrality bins are shown in Table~\ref{tab:cent_TAA}.
The STAR collaboration reports somewhat larger charm cross section 
(1.3 $\pm$ 0.2 $\pm$ 0.4 mb per $NN$ collision) 
in $p + p$ and $d +$ Au collision at $\sqrt{s_{NN}}$ = 200~GeV 
\cite{StarCharm}. The next-to-leading order pQCD calculation
of charm cross section is 300 to 450 $\mu$b \cite{NLOvogt}.

\begin{table}[htbp]
\caption{\label{tab:cent_TAA} 
Centrality bin, number of $NN$ collisions, nuclear 
overlap function, charm cross section per $NN$ collision, and total 
charm multiplicity per $NN$ collision, in  $\sqrt{s_{NN}} = 200$\,GeV 
Au+Au reactions.
}
\begin{ruledtabular}\begin{tabular}{r@{--}lr@{$\pm$}lr@{$\pm$}lr@{$\pm$}l@{\hspace{0pt}}lr@{$\pm$}l@{\hspace{0pt}}l}
 \multicolumn{2}{c}{Cen-}      & 
 \multicolumn{2}{c}{$N_{coll}$} & 
 \multicolumn{2}{c}{$T_{AA}$}  & 
 \multicolumn{3}{c}{$\frac{1}{T_{AA}}\frac{dN_{c\overline{c}}}{dy}|_{y=0}$} & 
 \multicolumn{3}{c}{$N_{c\overline{c}}$/$T_{AA}$} \\
 \multicolumn{2}{c}{trality}     &    
 \multicolumn{2}{c}{}            & 
 \multicolumn{2}{c}{(mb$^{-1}$)} & 
 \multicolumn{3}{c}{($\mu$b)}    & 
 \multicolumn{3}{c}{($\mu$b)}    \\ \hline
\multicolumn{2}{c}{min. bias} &    258&25  & 6.14&0.45 &143 &13 &$\pm$36 &622 &57  &$\pm$160 \\
 0&10 \%                      &    955&94  & 22.8&1.6  &137 &21 &$\pm$35 &597 &93  &$\pm$156 \\
10&20 \%                      &    603&59  & 14.4&1.0  &137 &26 &$\pm$35 &596 &115 &$\pm$158 \\
20&40 \%                      &    297&31  & 7.07&0.58 &168 &27 &$\pm$45 &731 &117 &$\pm$199 \\
40&60 \%                      &     91&12  & 2.16&0.26 &193 &47 &$\pm$52 &841 &205 &$\pm$232 \\
60&92 \%                      &   14.5&4.0 & 0.35&0.10 &116 &87 &$\pm$43 &504 &378 &$\pm$190 \\


\end{tabular}\end{ruledtabular}
\end{table}

It should be noted that final-state effects only influence the
momentum distribution of charm; they have little or no effect on the
total open charm yield. Therefore, our results indicate $N_{coll}$ scaling
of the initial charm production, as expected for point-like pQCD processes.
pQCD calculations without charm quark energy loss
and hydrodynamic calculations assuming complete thermalization of
charm quarks predict very similar heavy flavor electron spectra for
$p_T < 2$\,GeV/$c$~\cite{Charmflow}.  Differentiating between
these opposite physical pictures is only possible for $p_T >
2.5$\,GeV/$c$, where statistics of the current analysis are limited.





In conclusion, we have measured single electrons from heavy flavor
decays in Au + Au collisions at $\sqrt{s_{NN}} =$ 200 GeV. We observe
that the centrality dependence of charm quark production is consistent
with $N_{coll}$ scaling, as expected for hard processes.
The much larger Au + Au data
set collected by PHENIX in the 2003-04 run 
will allow us more detailed exploration of medium
effects on heavy quark production, both through deviations of the
heavy flavor electron spectrum from $N_{coll}$ scaling, and also
through a measurement of charm quark flow.

We thank the staff of the Collider-Accelerator and Physics
Departments at BNL for their vital contributions.  We acknowledge
support from the Department of Energy and NSF (U.S.A.), MEXT and
JSPS (Japan), CNPq and FAPESP (Brazil), NSFC (China), CNRS-IN2P3
and CEA (France), BMBF, DAAD, and AvH (Germany), OTKA (Hungary), 
DAE and DST (India), ISF (Israel), KRF and CHEP (Korea),
RMIST, RAS, and RMAE, (Russia), VR and KAW (Sweden), U.S. CRDF 
for the FSU, US-Hungarian NSF-OTKA-MTA, and US-Israel BSF.


\def\Journal#1#2#3#4{{#1}{\bf #2}, #3 (#4)}
\def\IJMPA{{Int. J. Mod. Phys.}~{\bf A}}
\def\JPG{{J. Phys}~{\bf G}}
\def\NCA{Nuovo Cimento}
\def\NIM{Nucl. Instrum. Methods}
\def\NIMA{{Nucl. Instrum. Methods}~{\bf A}}
\def\NPA{{Nucl. Phys.}~{\bf A}}
\def\NPB{{Nucl. Phys.}~{\bf B}}
\def\PLB{Phys. Lett. B}
\def\PLC{Phys. Repts.\ }
\def\PRL{Phys. Rev. Lett.\ }
\def\PRD{Phys. Rev. D}
\def\PRC{Phys. Rev. C}
\def\ZPC{{Z. Phys.}~{\bf C}}

\end{document}